\documentclass[manuscript,nonacm]{acmart}

\graphicspath{ {images/} }
\usepackage{subcaption}
\usepackage{caption}
\usepackage{comment}
\usepackage{makecell}
\usepackage{hyperref}
\usepackage{hyperxmp}

\usepackage{enumitem}
\usepackage[framemethod=tikz]{mdframed}

\AtBeginDocument{%
  \providecommand\BibTeX{{%
    \normalfont B\kern-0.5em{\scshape i\kern-0.25em b}\kern-0.8em\TeX}}}

\begin{document}

\title[From Passersby to Placemaking]{From Passersby to Placemaking: Designing Autonomous Vehicle-Pedestrian Encounters for an Urban Shared Space}

\author{Yiyuan Wang*}
\email{yiyuan.wang@sydney.edu.au}
\orcid{0000-0003-2610-1283}
\affiliation{%
  \institution{Design Lab, Sydney School of
Architecture, Design and Planning, The University of Sydney}
  \city{Sydney}
  \state{NSW}
  \country{Australia}
}

\author{Martin Tomitsch}
\email{martin.tomitsch@uts.edu.au}
\orcid{0000-0003-1998-2975}
\affiliation{%
 \institution{Transdisciplinary School, The University of Technology Sydney}
  \city{Sydney}
  \state{NSW}
  \country{Australia}
}

\author{Marius Hoggenmüller}
\email{marius.hoggenmueller@sydney.edu.au}
\orcid{0000-0002-8893-5729}
\affiliation{
 \institution{Design Lab, Sydney School of
Architecture, Design and Planning, The University of Sydney}
  \city{Sydney}
  \state{NSW}
  \country{Australia}
}

\author{Senuri Wijenayake}
\email{senuri.wijenayake@rmit.edu.au}
\orcid{0000-0003-1985-0227}
\affiliation{%
 \institution{School of Computing Technologies, The Royal Melbourne Institute of Technology}
  \city{Melbourne}
  \state{VIC}
  \country{Australia}
}

\author{Wai Yan}
\email{wai.yan@sydney.edu.au}
\affiliation{%
  \institution{Design Lab, Sydney School of
Architecture, Design and Planning, The University of Sydney}
  \city{Sydney}
  \state{NSW}
  \country{Australia}
}

\author{Luke Hespanhol}
\email{luke.hespanhol@sydney.edu.au}
\orcid{0000-0003-0839-481X}
\affiliation{%
  \institution{Design Lab, Sydney School of
Architecture, Design and Planning, The University of Sydney}
  \city{Sydney}
  \state{NSW}
  \country{Australia}
}

\renewcommand{\shortauthors}{Wang et al.}

\begin{abstract}
Autonomous vehicles (AVs) tend to disrupt the atmosphere and pedestrian experience in urban shared spaces, undermining the focus of these spaces on people and placemaking. We investigate how external human-machine interfaces (eHMIs) supporting AV-pedestrian interaction can be extended to consider the characteristics of an urban shared space. Inspired by urban HCI, we devised three place-based eHMI designs that (i) enhance a conventional intent eHMI and (ii) exhibit content and physical integration with the space. In an evaluation study, 25 participants experienced the eHMIs in an immersive simulation of the space via virtual reality and shared their impressions through think-aloud, interviews, and questionnaires. Results showed that the place-based eHMIs had a notable effect on influencing the perception of AV interaction, including aspects like visual aesthetics and sense of reassurance, and on fostering a sense of place, such as social interactivity and the intentionality to coexist. In measuring qualities of pedestrian experience, we found that perceived safety significantly correlated with user experience and affect, including the attractiveness of eHMIs and feelings of pleasantness. The paper opens the avenue for exploring how eHMIs may contribute to the placemaking goals of pedestrian-centric spaces and improve the experience of people encountering AVs within these environments.
\end{abstract}



\maketitle

\section{Introduction}
Autonomous vehicles (AVs) will not only operate on vehicle-dominant roads but also increasingly drive into urban shared spaces that prioritise pedestrians~\cite{clarke2006shared}. Shared space is a recent urban planning approach designed to minimise the demarcation between pedestrians and vehicles by removing kerbs, road markings, traffic signs, and traffic signals and therefore encourage all road users (RUs) to share the same public space with equal rights \cite{moody2014shared,clarke2006shared}. While vehicles drive slowly and more attentively to surroundings~\cite{wang2022pedestrian,li2021autonomous}, recent deployments of AVs passing by urban shared spaces have shown that AVs notably influenced pedestrian perceptions and experiences~\cite{predhumeau2021pedestrian}, often negatively affecting their behaviours around AVs~\cite{eden2017road,rodriguez2017safety} or their engagement in recreational activities~\cite{wang2022pedestrian}. Unlike regular vehicular roads, the shared space approach is driven by the impetus of ``placemaking'', seeking to reclaim urban spaces for people rather than vehicles~\cite{karndacharuk2013analysis,clarke2006shared}. In general, placemaking is defined as \textit{``the process of creating quality places that people want to live, work, play and learn in''}~\cite{wyckoff2014definition}. In transport, the shared space approach imbues space with meaningful values to cultivate a vibrant and welcoming atmosphere, thereby enhancing space usage, such as through increased street activities and extended dwell times~\cite{karndacharuk2013analysis,moody2014shared}. Supporting this objective are often the aesthetic and spatial rearrangements of the physical layout, for example, removing vehicular lanes, road marks, and traffic lights and adding street furniture~\cite{moody2014shared,hammond2013attitudes}. Given these characteristics, AVs entering shared spaces should consider supporting the quality of these places, including aspects such as visual appeal, pedestrians' well-being, and social interactions.

With the advent of AVs, external human-machine interfaces (eHMIs) have emerged as a novel interface type in human-computer interaction (HCI) to support AV-pedestrian interaction in the absence of human drivers \cite{dey2020taming,tran2021review}. Typical eHMIs are designed as part of automotive lighting systems, with the purpose of coordinating actions between AVs and pedestrians, compensating for the social cues such as eye contact or hand gestures that drivers usually use to communicate their intentions~\cite{rasouli2019autonomous,mahadevan2018communicating}. With the development of media technology and the internet of things, eHMIs are also envisioned to provide new affordances, paradigms, and experiences for pedestrians in their urban living, ranging from crosswalks projected on the ground~\cite{nguyen2019designing,daimler2015mercedes}, to manoeuvre messages transmitted to wearable devices~\cite{prattico2021comparing,tabone2023augmented}, to alerts of approaching AVs connected to urban infrastructure such as poles~\cite{chauhan2023fostering} or curbstones~\cite{hollander2022take}.

However, previous work on designing eHMIs has yet to consider how these interactions influence pedestrians' perceptions and experiences of places. \citet{gustafson2001meanings} found that people attribute ``meanings of place'' by considering the characteristics and behaviours of other entities within a space, the built and the natural environment, and the interplay between them, such as the atmosphere formed and the opportunities for desirable experiences. Therefore, as eHMIs increasingly become integral components of AVs~\cite{ISO/TR,Mercedes}, they hold promise to impact the ways pedestrians evaluate, relate to, and return to place as AVs navigate through shared spaces~\cite{gustafson2001meanings}. Prior research has primarily examined ``intent-based'' eHMIs, exploring their role in communicating the intentions of AVs to pedestrians. To date, research has not explored a ``place-based'' approach to eHMI design, which would consider their influence on the quality of urban environments. In this paper, we propose place-based eHMIs as those designed with contextual considerations to support placemaking in shared spaces. In our study, this includes how the eHMIs physically integrate with the environment (emitted, projected, or embedded) and how their aesthetics are conceived based on content that can potentially resonate with the users of the space.

In our exploration of place-based eHMIs, we aim to answer two research questions: (RQ1) \emph{How can eHMIs be designed to exhibit placemaking considerations for an urban shared space, i.e., place-based eHMIs?} (RQ2) \emph{How do place-based eHMIs influence pedestrian perception and experience of AV encounters in shared spaces?}

To investigate RQ1, we employ a Research-through-Design (RtD) approach to generate knowledge through the process of creating prototypes of place-based eHMIs, utilising its advantage of inquiring into exploratory, under-constrained problem spaces~\cite{zimmerman2007research}. To account for placemaking considerations, we draw on urban HCI~\cite{Fischer2012} literature, focusing on the aesthetic and spatial alignment of digital layers into physical urban spaces~\cite{willett2016embedded,hoggenmueller2019self,Wouters2016a}. To address contextual specifics and the relevance of designs to place~\cite{Dalsgaard2010,vande2012role}, our inquiry focuses on a real-world pedestrian zone local to where this research was conducted, where passenger pods with automation capability occasionally traverse. To investigate RQ2, we conduct an evaluation study where participants experience prototypes of the place-based eHMIs via virtual reality (VR) created from a 360-degree real-world footage of the pedestrian zone. Through think-aloud protocols, semi-structured interviews, and questionnaires, we identify effects of place-based eHMIs on influencing how participants perceive and experience the shared space and their AV encounter within this space.

\enlargethispage{2\baselineskip}

\section{Related Work}

\subsection{External Human-Machine Interfaces}
As human drivers delegate control to AVs, interaction design research has increasingly focused on exploring eHMIs to facilitate interactions between AVs and pedestrians~\cite{mahadevan2019av,dey2020taming,tran2021review}. These eHMIs are important for enabling pedestrians to comprehend the intentions and maneuvers of AVs, effectively replacing traditional driver-pedestrian communication channels such as eye contact and hand gestures~\cite{mahadevan2018communicating,rasouli2019autonomous}. A growing body of design efforts dedicate to envisioning pedestrian life in future urban environments; for example, AVs have been designed to utilise a broad range of digital and display technologies, such as laser projection~\cite{nguyen2019designing,daimler2015mercedes} or augmented reality imagery~\cite{prattico2021comparing,tabone2023augmented}, to communicate their messages; as the traffic system evolves into an intelligent, interconnected network, studies have also proposed AVs to communicate with pedestrians through common urban infrastructure~\cite{hollander2022take,chauhan2023fostering} or pedestrian-carried devices~\cite{mahadevan2018communicating,tran2022designing}.

Research studies have investigated the usability aspects of eHMIs, for example, it is found that a cyan coloured, uniformly pulsating LED light band attached to the front of AV is effective for suggesting a yielding message (i.e., AV stopping and giving way)~\cite{dey2020color} and for supporting pedestrian crossing behaviours at uncontrolled roads~\cite{colley2022effects,lanzer2023interaction}. Furthermore, with the ``waves'' of HCI focus shifting from valuing mere technology efficiency to the quality of experience throughout the technology integration \cite{lopes2021hci}, eHMI studies have investigated qualities of pedestrian experience in AV interaction, such as the user experience (UX) of eHMIs and the perceived safety of AV interaction~\cite{locken2019investigating,colley2022effects,faas2020longitudinal}. Additionally, there have been designs that consider the impact on emotional or social aspects; for instance, emotional expressions on AVs could improve the sociability of AVs in pedestrianised zones~\cite{wang2023my,wang2023designing}, a perceived malfunction of eHMIs could lead to negative pedestrian emotions~\cite{m2021calibrating}, and eHMIs that convey prosociality (i.e. mindfulness and care for pedestrians)~\cite{sadeghian2020exploration} or positive social feedback~\cite{colley2021investigating} could result in improved ``traffic climate''~\cite{sadeghian2020exploration} or the social acceptability of AVs~\cite{colley2021investigating}.

Furthermore, as pervasive displays increasingly move towards distributed and autonomous deployments~\cite{claes2018conveying}, there have been creative utilisations of vehicles that spur changes to their traditional roles in cities, from trams adorned with festive LED lights, e.g., ``urban pixels''~\cite{seitinger2009urban} to the more recent repurposing of the exteriors of AVs into public displays~\cite{Colley2018,asha2020designing}. These designs demonstrate new possibilities of inspiration and affordances that eHMIs may offer to pedestrians within urban environments.

\subsection{Urban HCI}
In the wake of technological advances, HCI researchers increasingly began to study interactive systems in contexts outside of workspaces~\cite{Bodker2006}, including the contemporary city~\cite{tomitsch2017making}. Broadening the context of use has also shifted the research focus from mere usability questions to considering the experiential qualities of interaction~\cite{lopes2021hci}. Designing technologies for everyday urban life further comes with unique design challenges that require the integration of knowledge from fields such as urban planning, architecture, and social sciences, in addition to HCI's disciplinary traditions. Under the umbrella term urban HCI~\cite{Fischer2012}, research has investigated the spatial, social, and cultural aspects in the design of interfaces that connect people, place, and technology. These interfaces may cut across multiple scales of the city \cite{gardner2018smlxl}, from individuals and crowds to precinct-wide interventions. Crucially, they address behaviour unfolding in public spaces, therefore subscribing to distinct tacit norms \cite{goffman2002presentation} for interactions between people and the surrounding built, natural, and digital environments. In doing so, they effectively augment physical urban spaces with dynamic digital layers \cite{hespanhol2022augmented}, in the process transforming the perceptions, affordances, and ``sense of place''. Examples range from place-based mixed reality \cite{innocent2016play, sanaeipoor2020smart}, public displays, and urban screens, to large-scale media fa\c{c}ades~\cite{Tscherteu2011}. They represent a digital strategy to placemaking with the promise to enhance the quality of urban experiences and to foster meaningful interactions~\cite{hespanhol2017media, quteprints88894}. 

However, researchers have also repeatedly reported on the failed deployment of urban technologies and interfaces, causing emotional disconnect between people and place, manifested, amongst others, through various forms of vandalism~\cite{Moere2012a, Dalsgaard2010}. These interventions failed to consider contextual aspects in regards to 
alignment of content and spatial integration of display technology into existing structures. Regarding the latter, researchers documented different approaches to the successful integration of display technology into the surrounding physical environment~\cite{hoggenmueller2019self, Wouters2016a,willett2016embedded}, such as using projection or light-emitting technologies. These have the advantage to blend with physical structures, with the digital layer perceived as an integral part of its carrier rather than an attached layer on top.

More recently, researchers in urban HCI have also turned to robots and AVs as a placemaking strategy~\cite{Colley2018}. For example, equipped with a low-resolution lighting display (for external communication) and capable to draw on the ground, a small-scale urban robot~\cite{hoggenmueller2020stop} successfully attracted passers-by and reactivated an abandoned laneway. While designed for playful interactions, such case studies can serve as inspiration on how eHMI designs in more mundane AV applications could preserve and unlock desirable qualities of city life~\cite{Nagenborg2018}.

\section{Methodology}
As the goal of this work is to explore the impact of eHMIs on how pedestrians perceive and experience places and AVs traversing through those places, we apply an RtD approach to acquire insights by designing eHMI prototypes that can assist us in understanding this under-constrained problem space~\cite{zimmerman2007research}. Due to the current lack of theoretical guidance on how to select from or experimentally manipulate existing intent eHMIs for placemaking considerations, we created speculative, bespoke place-based eHMI designs through the RtD process, extending a conventional intent eHMI. Pertaining to this, ~\citet{gaver2012should} argues for the generative power of design artefacts and how they can embody conceptual insights from the RtD process. To that end, we describe our design considerations, decisions and implementation of the eHMI prototypes that seek to influence pedestrian perceptions and experiences within urban places, thereby giving dimensionality to the design space~\cite{gaver2012should}.

\subsection{Designing eHMIs for a Local Shared Space}

Urban HCI highlights the significance of contextual aspects when designing for interfaces between people and interactive systems in public spaces~\cite{Fischer2012,gardner2018smlxl, Dalsgaard2010}. We therefore anchored our design to a local shared space (see Figure~\ref{shared-space}) -- an urban corridor frequented by pedestrians and occasionally passenger pods, including test AVs developed by our university's engineering faculty. We further centered our design on a natural and daily encounter where a pedestrian and an AV casually approach each other, as this scenario was identified as the most frequent form of naturalistic vehicle-pedestrian interaction by our prior longitudinal observation at this location~\cite{wang2022pedestrian}. Based on this setting, we derived our design considerations and decisions, presented in sections below.

\subsubsection{Design Considerations}
\begin{figure*}[htbp]
    \centering
    \begin{subfigure}[b]{0.4\textwidth}
        \centering
        \includegraphics[width=\textwidth]{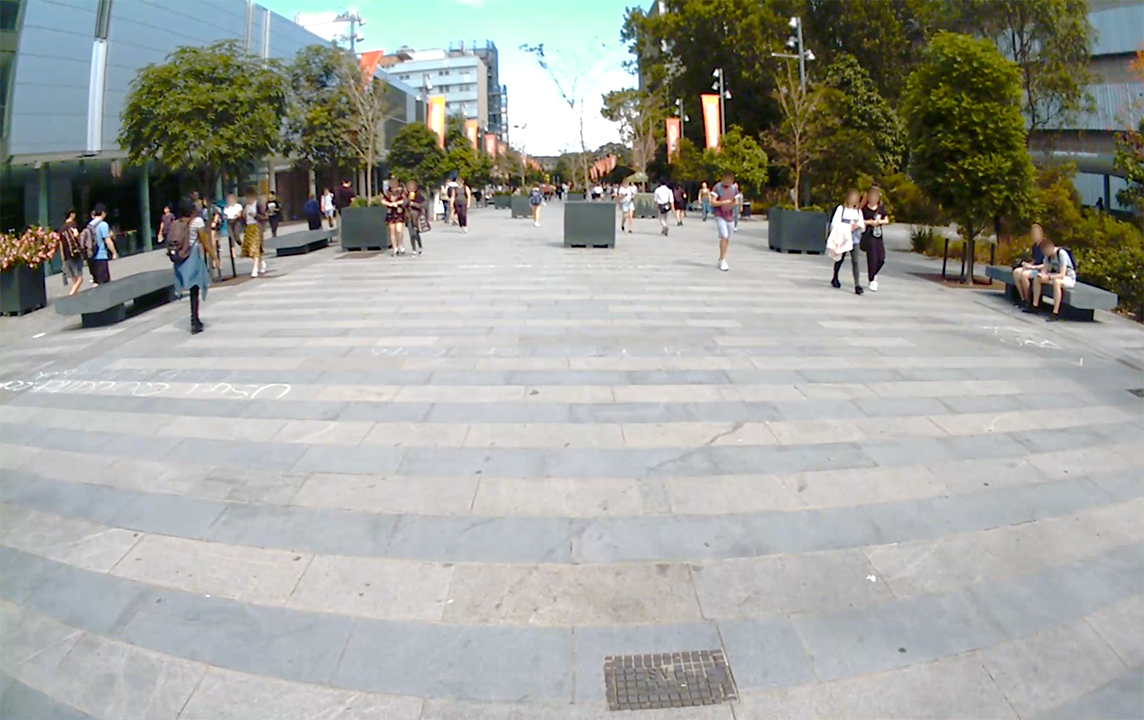}
        \caption{}
        \label{shared-space}
    \end{subfigure}
    \hfill
    \begin{subfigure}[b]{0.59\textwidth}
        \centering
        \includegraphics[width=\textwidth]{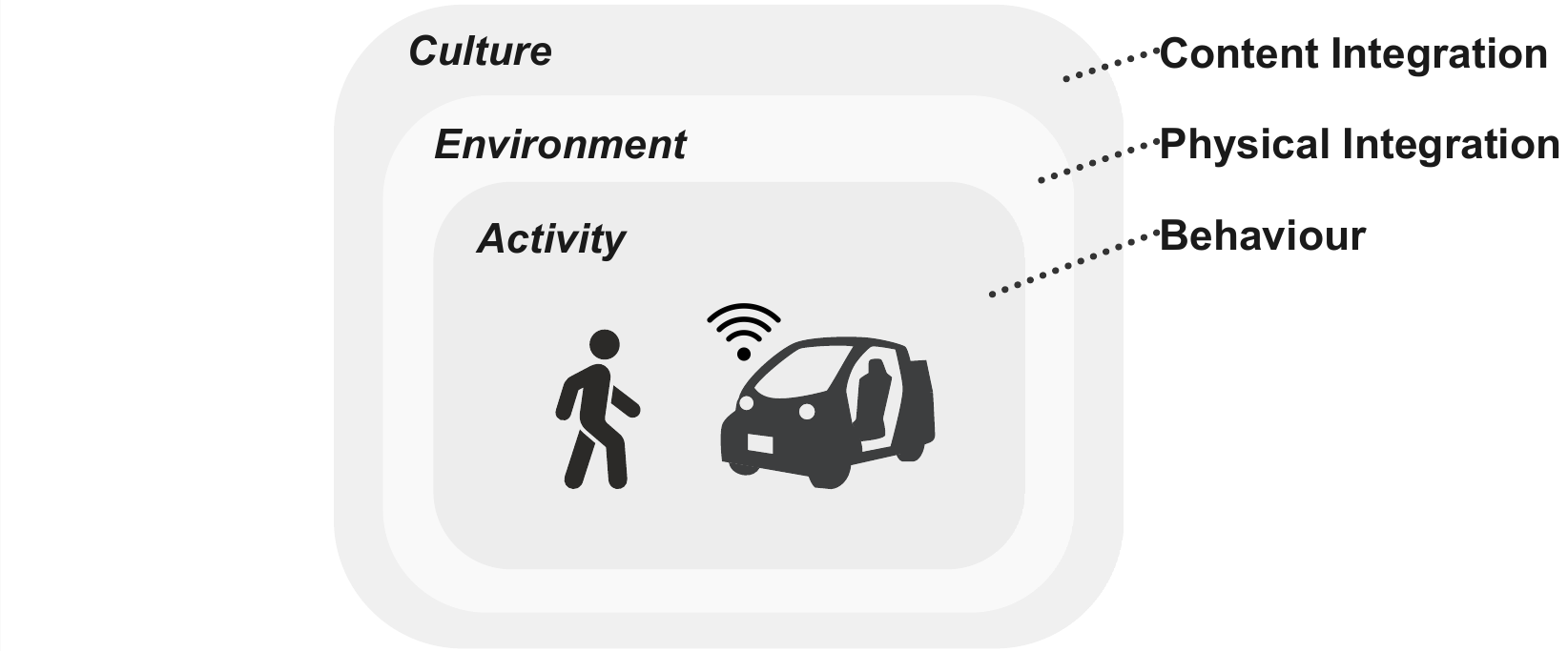}
        \caption{}
        \label{mapping}
    \end{subfigure}
    \caption{The local shared space (a). Designing place-based eHMIs needs to consider multiple layers of contexts: the pedestrian's activity, the environment, and the culture, dimensions provided by \citet[p. 60]{tomitsch2017making} for designing smart urban HCI interventions (b).}
\end{figure*}

We based our design considerations on a framework of contextual dimensions for designing smart urban HCI interventions~\cite{tomitsch2017making} (~\autoref{mapping}). The framework highlights the multiple layers of contexts that designing interfaces within urban public environments should consider, including the user's situated activity (activity), the physical, built environment (environment), and the socio-cultural dimension which the interface operates within (culture). We therefore derived three considerations for place-based eHMIs, namely behaviour, physical integration, and content integration, addressing the AV-pedestrian interaction, the physical environment, and the socio-cultural context.

\raggedbottom
 
\textbf{Behaviour} Based on recent trends in policies and standards~\cite{Mercedes,ISO/TR}, future AVs are likely to be equipped with utilitarian eHMIs that communicate messages related to intent or status. Following this trend, we decided to design eHMIs that align with the functional message conveyed by the AV. Hence, the behaviour of our eHMIs was based on a cyan coloured, sinusoidally pulsating LED light band attached to the front of AV that signals a stopping intent. This simple light-band eHMI was found by~\citet{dey2020color} in a crowdsourced survey to most effectively communicate the stopping intent in a non-contextualised setting. Recent studies have used this light-band eHMI to compare with alternative eHMIs and to study pedestrian behaviours~\cite{colley2022effects,lanzer2023interaction}.

\textbf{Physical Integration} To design for the effect of eHMIs on place, we turned to practices from urban HCI that integrate the digital layer of technology to existing structures in urban public spaces~\cite{hoggenmueller2019self, Wouters2016a,willett2016embedded}, serving to connect people with digital technology and with the surrounding built environments~\cite{schroeter2012people,goffman2002presentation,gardner2018smlxl}. In this light, we first considered the physical dimension of how eHMIs can interplay with the shared space, guided by two questions: (i) what mechanisms, as documented in urban HCI literature, can facilitate the integration of digital technology with physical structures? (ii) what are conceivable modalities of eHMIs, as supported by current vehicle technology or eHMI literature, that can synergise with those integration mechanisms? Based on related frameworks in urban HCI~\cite{hoggenmueller2019self,Wouters2016a,willett2016embedded}, we derived three eHMI modalities that demonstrate varying mechanisms of integration with the physical environment: 
\begin{itemize}
    \item Emitted: light emitted from the AV affects the luminosity of the immediate surroundings; 
    \item Projected: the AV projects content onto the road, applying a transient fabric to the physical surface; 
    \item Embedded: the physical structure is digitally activated to respond to the AV.
\end{itemize}
These mechanisms also aligned with conceivable and desirable vehicle technology, as supported by industry efforts~\cite{smartRoad2017,bmweink,daimler2015mercedes} and previous eHMI studies~\cite{nguyen2019designing,hollander2022take}.

\textbf{Content Integration} We considered the importance to align the content depicted by eHMIs with the socio-cultural specifics of the chosen locale~\cite{Dalsgaard2010,willett2016embedded}. This included our reflections on the values that this shared space provides for pedestrians, for example, not only commute but also recreation, as well as what content might resonate with people who use or pass by the place. Motivated by the increasing body of placemaking practices that use digital representations of nature elements as solutions to foster civic engagement and well-being in built environments~\cite{de2023digital,hespanhol2022augmented}, we decided to guide the ideation process by drawing inspirations from nature. Since the socio-cultural aspects of interface design are likely to be location-specific and difficult to be directed by a blueprint approach~\cite[p. 62]{tomitsch2017making}, we conducted an ideation workshop to inform the content of our final design concepts.

\begin{figure}[h]
  \centering
  \includegraphics[width=0.9\textwidth]{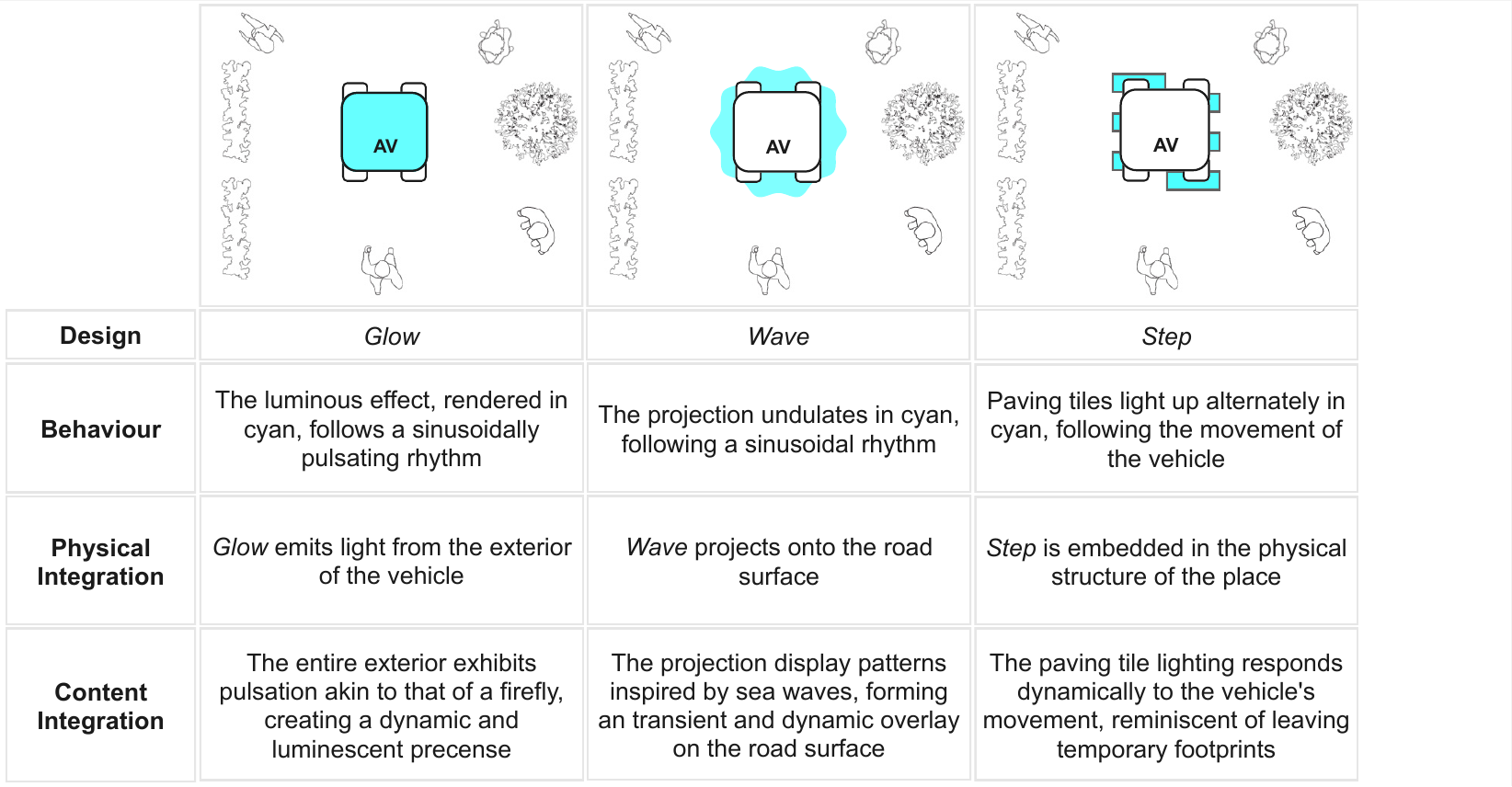}
  \caption{Descriptions of the characteristics of our place-based eHMI designs: \emph{glow}, \emph{wave}, and \emph{step}}
  \label{design}
\end{figure}

\subsubsection{Ideation Workshop}
We conducted an interdisciplinary workshop convening ten researchers from across interaction design, HCI, media architecture, and vehicle automation engineering. The workshop was facilitated by the paper's first author and included two other authors. Workshop participants were frequent users of the chosen shared space, allowing them to bring their knowledge of the space and personal experience to the ideation process along with their domain knowledge. The workshop was held online using video conferencing and a collaborative digital whiteboard. In the beginning, participants collectively revisited the characteristics of this shared space (10 minutes), including its functionality and daily ambience. Subsequently, participants were divided into three groups, each facilitated by one author, to engage in open discussions (10 minutes), after which the facilitator of each group presented key points to everyone (5 minutes each, 15 minutes total). The group discussions were driven by the question: \emph{how can nature inspirations support AVs to interact with pedestrians in this shared space?} with a focus on the scenario where the AV and the pedestrian casually approach each other. Finally, participants engaged in a collective discussion based on the presentations of group discussions (15 minutes).

We collected the following main points raised in the group discussions: (i) the ability to reach a broad audience, e.g., by avoiding inspirations that might evoke fear in some pedestrians; (ii) how information was signalled in nature, such as a mention of using ``changes in colour'' to communicate intentions; (iii) how nature elements could interplay with the mobility of AVs, as AVs could have a ``fluid'' influence on their surroundings as navigating through them. The latter stimulated a lively discussion among all participants. Participants noted using nature elements, e.g., marine-related, to provide connections between qualities of ``fluidity'' and the mobility of AVs, thereby rendering AVs as naturally blending and engaging with the surroundings as they pass by this shared space.

\subsubsection{Design Concepts}
In light of these discussions, the design team ideated content in relation to each of the three physical integration mechanisms. Firstly, for the emitted mechanism, we considered participants' suggestion of applying luminescence to the ``skin'' of the AV. We also needed to think about how to convey its behaviour, which was to synchronise with the uniform pulsation of the light-band intent eHMI. Consequently, we captured the essence of utilising the exterior of the AV, allowing it to pulsate in the same colour and speed as the light band. Next, in designing the projected mechanism, we recognised the transient nature of projection, i.e., temporarily manipulating the fabric of a physical surface. We then drew parallels with a local natural phenomenon of bioluminated sea waves, echoing the visual effect of applying a momentary layer to the sea shore. To design for its behaviour, we made the projected waves to expand and contract at the same rate as the light band. Finally, when devising the embedded mechanism, we reflected on participants' mentions of certain marine creatures that can ``leave an ink trail'' to interact with the physical environment. Drawing on this cue, we imagined an AV leaving a trace in the environment, such as the paving tiles in this shared space reacting to the AV's movement. We also considered how long such responsiveness from the tiles should sustain, for example, aligning them more closely with the ``in-situ'' movement of the AV rather than leaving a prolonged trace. We finally arrived at three designs, namely \emph{glow}, \emph{wave}, and \emph{step}, depicting varying characteristics concerning the behaviour, physical integration, and content integration of eHMIs in relation to the shared space, as detailed in~\autoref{design}.

\subsection{Implementing eHMI Prototypes in VR}
VR simulations are recognised for their ability to assess exploratory eHMI proposals~\cite{tran2021review,mahadevan2019av} and to guarantee pedestrian safety during interactions with AVs~\cite{deb2017efficacy,nascimento2019role}. In recent years, more AV studies have turned to 360-degree real-world videos to create immersive, realistic traffic environments in VR~\cite{yeo2020toward,hoggenmuller2021context,chang2022can,wang2024immersive} for their ability to capture omnidirectional on-site information in high visual-audio fidelity~\cite{hoggenmuller2021context,yeo2020toward,vettehen2019taking,wang2024immersive}. Therefore, to set up a realistic representation of the shared space in VR, we recorded 360-degree panoramic videos of the space using 8K Insta360 Pro as our video and sound recording device. After post-processing to reduce image noise, we imported the video into Unity and created an immersive VR environment, with the original ambient noise retained as soundscape.

In supporting the implementation of eHMIs, we created a high-fidelity 3D model of an autonomous passenger pod which at the time of the study operated in this shared space for collecting driving data. The vehicle was modelled in Autodesk 3ds Max and imported into Unity for VR development. To implement \emph{glow}, we applied an emissive material as texture to the exterior of the AV and changed its visual properties to simulate the pulsating effect. A particle system from the physics library was utilised to simulate the bioluminescent projection of \emph{wave}. To implement \emph{step}, we created virtual overlays of the physical paving tiles and animated them to light up and disappear as the AV rolled over them.

As the behaviour of the place-based eHMIs was to amplify a conventional intent eHMI, i.e., a pulsating light band indicating stopping, based on work by~\citet{dey2020color}, we added the light band as a \emph{baseline} design. Additionally, a synthetic, fading engine sound was incorporated to augment the deceleration of the AV, offering a visual-audio conveyance of intent. The light band and the engine sound were both applied to \emph{glow}, \emph{wave}, and \emph{step} to support comparability. Additionally, we adjusted the colour of \emph{glow} from cyan to blue, as we found from our pilot study with 4 people, that it might cause difficulty to discern the light band from the exterior of the same colour, undermining the effect of the light band. The final prototypes were deployed to Oculus Quest 2 for user evaluation.

\subsection{Evaluation Study}

\begin{figure}[h!]
  \centering
  \includegraphics[width=\textwidth]{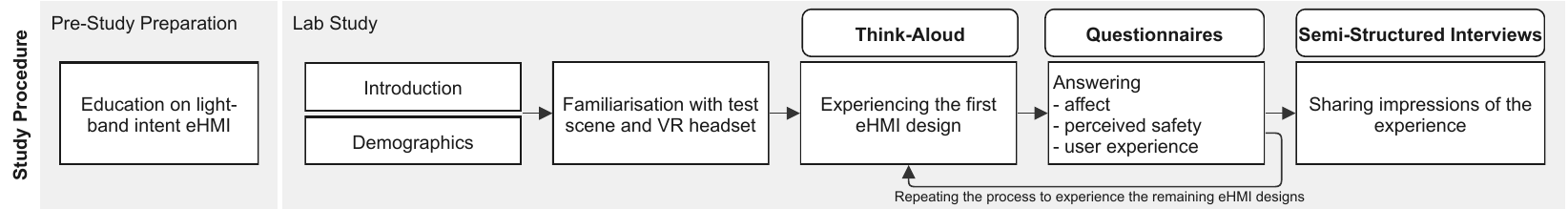}
  \caption{An overview of the procedure of evaluation study.}
  \label{procedure}
\end{figure}

\subsubsection{Study Procedure}
An overview of the study procedure is illustrated in~\autoref{procedure}. As future AVs are increasingly expected to be equipped with intent eHMIs~\cite{faas2020longitudinal,ISO/TR,Mercedes} and the light band serves as a popular representation~\cite{dey2020color,colley2022effects,lanzer2023interaction}, we provided participants with preliminary education on the light band and its stopping intention. This would also help us gather responses that are more relevant to our proposed designs, since the light band was incorporated into all three place-based designs. Three days before each individual session, a one-page slide explaining the light-band eHMI was provided to each participant via email. The slide contained (i) a cyan, sinusoidally pulsating light band (animated image), and (ii) texts reading \emph{``A flashing light band attached to the lower front of an autonomous vehicle indicates the vehicle will stop and give way to you''}. Participants were required to remember this information before coming to the session. The slide was again provided via email one day before the session and on-site at the beginning of the session.

\begin{figure}[h]
  \centering
  \includegraphics[width=\textwidth]{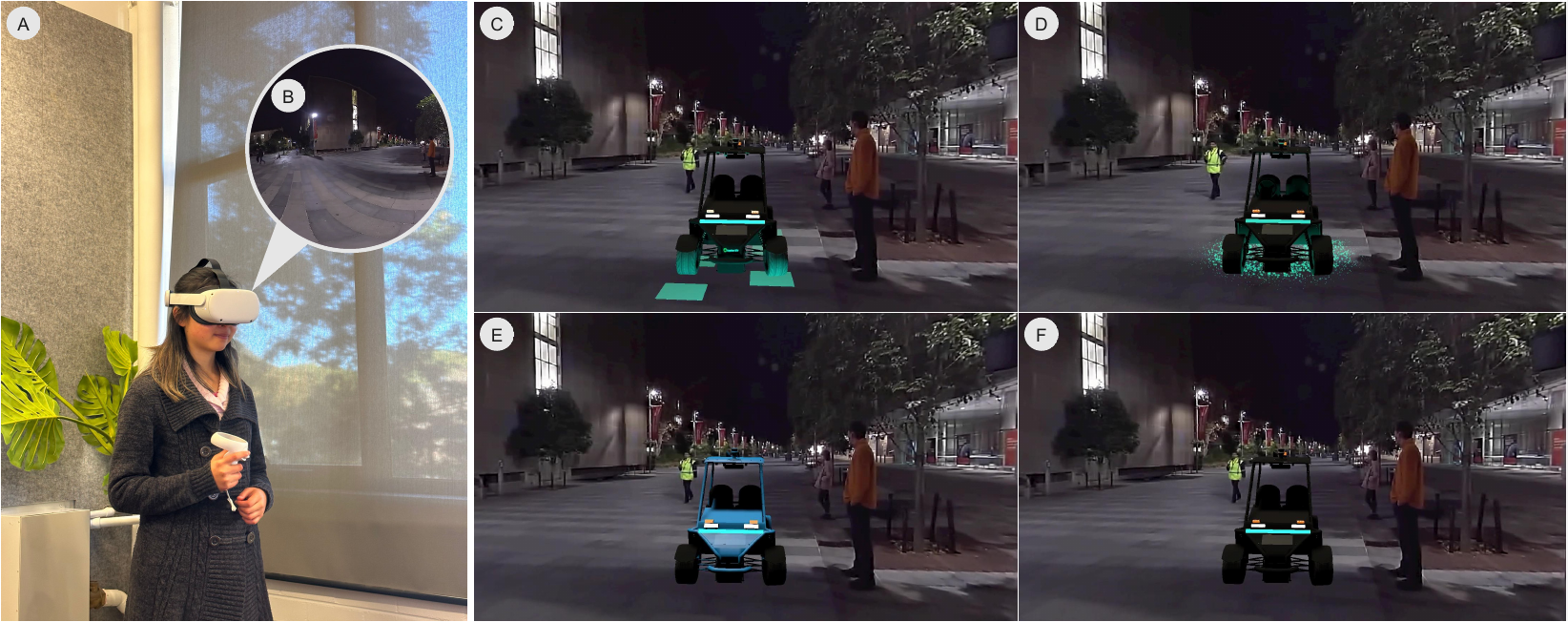}
  \caption{A participant experiencing the prototypes through a VR headset (A); the 360-degree video-based environment in VR (B); place-based eHMI prototype implementations of the \emph{step} (C), \emph{wave} (D), and \emph{glow} (E) concepts, and the \emph{baseline} eHMI (F).}
  \label{prototypes}
\end{figure}

After receiving consent from participants, we provided a brief overview of the research topic and the tasks included in the evaluation. We explained the shared space approach to participants and prescribed the following scenario to them: \emph{``You are walking in this shared space and want to go down the street''}. Then, we asked participants to put on the VR headset and immerse themselves in a test scene of the shared space environment. After familiarising with study equipment and environment, participants experienced the three designs and the baseline (see~\autoref{prototypes} for eHMI prototypes) in an order randomised with a balanced Latin Square (LS) to minimise carryover effects. With four conditions, the first 24 participants were evenly counterbalanced, and the 25th participant was randomly assigned to one of the orders. For each design, the AV drove at a slow speed of 2 m/s and decelerated to a complete stop at a distance of about 2 m from the participant's location. Participants were asked to think aloud while encountering the AV. Before participants filled in the questionnaires for the first time, we explained the content and instructions to ensure they understood how to complete them. After each design experience, participants completed a questionnaire before moving on to the next design. At the end of the session, participants took part in a semi-structured interview.

\subsubsection{Participants}
We recruited 25 participants (10 male, 15 female; age range of 20-50 years: M=28.68, SD=6.56) through physical flyers and social networks. All participants were required to have normal or corrected-to-normal eyesight and speak fluent English. Twenty participants frequently used the shared space, and five participants had only been there a few times. Fourteen participants' main transport mode was public transport (e.g., train, bus), while six primarily travelled by car, four walked, and one cycled. The study took an average of 51.76 minutes (SD=9.24) to complete, which was estimated at 60 minutes based on the pilot study with 4 people. All study sessions were audio-recorded. Participants were each compensated with a \$40 gift voucher, following the study protocol approved by the Human Research Ethics Committee at the University of Sydney.

\subsection{Data Collection and Analysis}

\subsubsection{Qualitative Data}
\label{interview}
We collected qualitative data in the form of think-aloud and semi-structured interviews. For think-aloud, participants were asked to spontaneously verbalise any immediate thoughts or feelings during their encounter with the AV. A semi-structured interview was conducted at the end of each study session, with the guidance of the following questions: (1) \emph{How would you rank the four designs based on your preference?}, (2) \emph{How do you usually feel in a shared space?}, and (3) \emph{How do you think the designs impacted or will impact your experience in the shared space?}.

We transcribed audio recordings of think-aloud and semi-structured interviews with the aid of a professional AI-supported transcription service \footnote{https://otter.ai/} and manually reviewed the transcripts. This process resulted in a total of approximately 53000 words of transcripts with an average length of 2130 words per participant. Guided by RQ2, two coders collaboratively analysed the data following an inductive thematic analysis approach~\cite{braun2006using}. First, the two coders worked on the same subset of data (one-third) and independently developed sub-themes. Afterwards, the two coders discussed and agreed on the sub-themes, followed by independently analysing the rest of the data. Finally, the two coders convened again to group sub-themes into final themes.

\subsubsection{Quantitative Data}
\label{questionnaire}
We assessed qualities of experience in the AV encounter from aspects of UX, affect, and perceived safety. The UX of eHMIs was measured with UEQ questionnaire \cite{laugwitz2008construction}. The questionnaire consists of 26 semantic differentials (7-points, -3 to +3) evaluating 6 subscales: \emph{attractiveness} (overall impression of the product), \emph{efficiency} (solving tasks without unnecessary effort), \emph{perspicuity} (how easy it is to get familiar with the product), \emph{dependability} (feeling in control), \emph{stimulation} (how exciting and motivating it is to use the product), \emph{novelty} (how innovative and creative the product is). Pragmatic qualities include \emph{efficiency}, \emph{perspicuity}, \emph{dependability}, and hedonic qualities include \emph{stimulation} and \emph{novelty}. The affect of pedestrians during interaction was measured with 9-point Self-Assessment Manikin (SAM)~\cite{bradley1994measuring} consisting of \emph{valence} (positive or negative connotation of emotion) \emph{arousal} (intensity of emotion) and \emph{dominance} (controlling or submissive nature of emotion). The perceived safety of the AV system was measured with an adapted Godspeed questionnaire \cite{bartneck2009measurement} by recent eHMI studies \cite{faas2020longitudinal,colley2022effects}. The adapted questionnaire contains 4 semantic differentials (7-points, -3 to +3): anxious/relaxed, agitated/calm, unsafe/safe, timid/confident.

We assessed the internal reliability (Cronbach's alpha) for questionnaires requiring score aggregation, that is, the UEQ questionnaire and the adapted perceived safety scales. Based on descriptors from literature \cite{taber2018use}, the overall internal reliability was strong for \emph{attractiveness} ($\alpha$=0.92) and \emph{perceived safety} ($\alpha$=0.94); good for \emph{perspicuity} ($\alpha$=0.87), \emph{dependability} ($\alpha$=0.80), \emph{stimulation} ($\alpha$=0.80), and \emph{novelty} ($\alpha$=0.70); and adequate for \emph{efficiency} ($\alpha$=0.67). Then, we examined the normality distribution of each of the above measures with Shapiro–Wilk tests and opted for Repeated Measures ANOVA with Greenhouse-Geisser corrections or Friedman tests (non-parametric) based on the normality of data. In case of significant differences, we performed post-hoc tests with Bonferroni correction for pairwise comparisons. Furthermore, Spearman's correlation coefficients were computed to assess the linear relationship between UX, affect, and perceived safety for each design.

\section{Results}
\begin{figure}[h]
  \centering
  \includegraphics[width=0.72\textwidth]{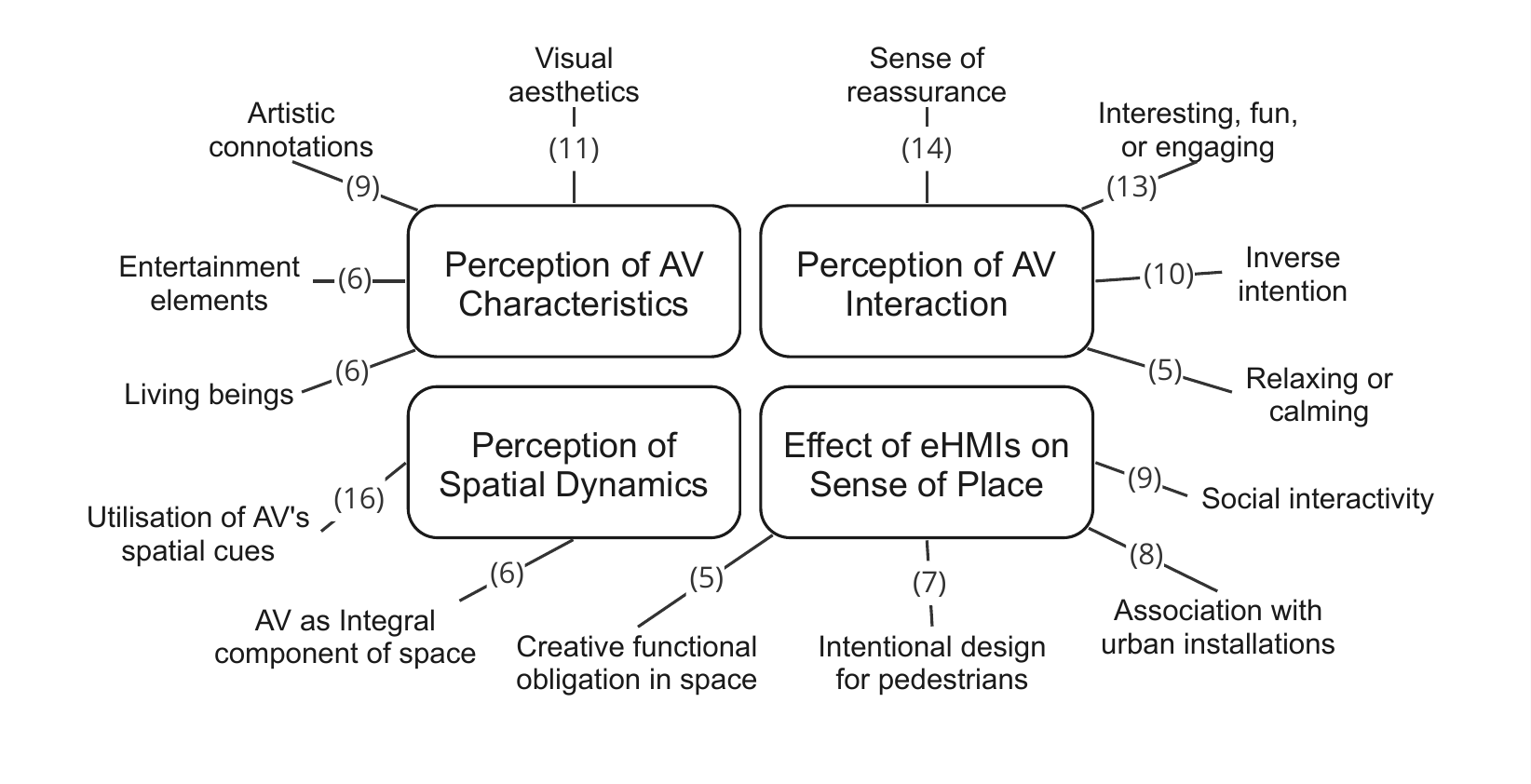}
  \caption{Four themes identified from participants' qualitative feedback, along with their sub-themes and the number of participants providing feedback relevant to the sub-theme.}
  \label{themes}
\end{figure}

Our qualitative analysis identified four themes presented in the following sections: (i) perception of AV characteristics; (ii) perception of AV interaction; (iii) perception of spatial dynamics; (iv) effect of eHMIs on sense of place. ~\autoref{themes} illustrates the sub-themes and their counts under these four themes. In the next four subsections, we elaborate on these sub-themes by quoting participants' individual considerations that qualitative analysis found to support these themes. Additionally, \autoref{correlation} presents quantitative results from questionnaire responses to UX, affect, and perceived safety and reports on correlations between these experiential qualities.

\raggedbottom

\enlargethispage{2\baselineskip}

\subsection{Perception of AV Characteristics}
Participants drew various associations upon encountering the AV equipped with the eHMIs. Eleven participants recognised the visual aesthetics of these interfaces, with nine participants suggesting artistic connotations that these interfaces embodied. Furthermore, for some participants, the designs were considered ornamental, more of ``an aesthetic advancement rather than a technical advancement'' (P8) or ``more performative than something for utility'' (P3).

\vspace{1em}
\hfill\begin{minipage}{\dimexpr\textwidth-1cm}
P8 (think-aloud, referring to \emph{wave}): ``This is not something which you usually see in a vehicle. So it kind of makes me think that it's an art piece or a creative thing rather than a normal thing.''

P10 (interview, referring to \emph{glow}): ``This will be a very cool [lighting] installation. Imagine a car just around the place with all this stuff -- I think it will be very aesthetic.''

\xdef\tpd{\the\prevdepth}
\end{minipage}
\vspace{1em}

Equipped with these interfaces, the AV was associated with living beings by six participants, for example, described as ``an organism growing and expanding'' (P8) or moving ``step by step like how we walk'' (P1). 
A possible explanation for these perceptions is the interplay between the intrinsic mobility of vehicles and the nature-inspired designs, imbuing the AV with a sense of vitality and even agency:

\vspace{1em}
\hfill\begin{minipage}{\dimexpr\textwidth-1cm}

P9 (interview, referring to \emph{wave}): ``This feels organic and smooth, and that's why it's engaging. As it becomes larger and smaller, it reminds me of breathing.''

P16 (think-aloud, referring to \emph{wave}): ``It's pulsating almost like a heartbeat. I think that makes it more -- I don't want to say relatable -- I guess it's more pleasing. It's got a human element to it.''

P20 (think-aloud, referring to \emph{glow}): ``It feels like it's really asking me politely like `Excuse me, please move'. The other times I didn't feel that it was trying to ask me something. I don't know why it's giving me polite vibes. I think, similar to that interaction where you're walking on the sidewalk and you're going to come to meet someone, you end up like doing that awkward pause with the stranger.''

\xdef\tpd{\the\prevdepth}
\end{minipage}
\vspace{1em}

Additionally, six participants perceived elements of entertainment, drawing references that they resonated as part of their personal experiences, ranging from well-received movies to familiar video games:

\vspace{1em}
\hfill\begin{minipage}{\dimexpr\textwidth-1cm}

P13 (interview, referring to \emph{step}): ``Everything's in a square, like it's gamified. I find that interesting to look at as well...like when you're playing Mario Kart. It reminds me of the scenes when you're stepping on something.''

P16 (interview, comparing \emph{step} to \emph{wave}): ``It reminds me of Saturday Night Fever -- kind of looks like a dance floor. It's a bit old school, reminds me of my kitchen floor when I was a child, like black and white...whereas the lighting of [wave] is more modern in terms of its approach.''

\xdef\tpd{\the\prevdepth}
\end{minipage}

\subsection{Perception of AV Interaction}

One of the main considerations behind our designs was to echo the functional message, i.e., a stopping intent, conveyed by the AV. Aware that the AV would stop, as they had learned previously that the pulsating light band signifies yielding, fourteen participants indicated a sense of reassurance, mentioning descriptors such as ``secure'' or ``confident'' because of the proposed designs reinforcing the stopping message:

\vspace{1em}
\hfill\begin{minipage}{\dimexpr\textwidth-1cm}
P4 (interview, referring to \emph{glow}): ``When the whole vehicle was flashing, it gave me a very dominant, not dominant, but obvious way of showing that it's going to stop, and that made me feeling a little bit more safe, compared to others.''

P18 (interview, referring to \emph{step}): ``I think the lights down the car were informative as it was telling me that there was more of a process as to what it was doing, so it would give me more confidence in knowing that it's about to stop, which was visually interpreted with the lights down the car.''

\xdef\tpd{\the\prevdepth}
\end{minipage}
\vspace{1em}

However, these enhancements could sometimes be perceived as assertive, leading some participants to feel less in control and to interpret the message contrarily, as if the vehicle was requesting passage. This sentiment was mostly associated with \emph{glow}, by eight participants, as its illumination was often linked to emergency situations; similarly, \emph{wave} and \emph{step} were each associated by one participant with ``police car flashing'' (P14) and ``power walking'' (P20) respectively, evoking a sense of compulsion.

\vspace{1em}
\hfill\begin{minipage}{\dimexpr\textwidth-1cm}
P5 (think-aloud, referring to \emph{glow}): ``There's a bit more of an urgency to this lighting that would make me want to move out of its way. It seems to be more similar to emergency vehicles that I'm more used to in real life.''

P15 (think-aloud, referring to \emph{glow}): ``I think the area of the colour is really big and it's a little bit more confronting...Now it's getting more closer. I feel like it's wanting me to give way to it.''

\xdef\tpd{\the\prevdepth}
\end{minipage}
\vspace{1em}

Regardless of the influence in coordinating actions, the proposed designs were described as interesting, fun, and engaging by thirteen participants. This aspect not only made pedestrians ``more interested in paying attention to it'' (P21), but also increased a sense of involvement into the AV's interaction:

\vspace{1em}
\hfill\begin{minipage}{\dimexpr\textwidth-1cm}
P8 (interview, referring to \emph{wave} and \emph{glow}): ``In those certain signage, it feels personal, like it's meant for the person who is moving on the tile or for the person who is dancing together. It's indicating something, and that indication is not general.''

P9 (interview, referring to \emph{wave} and \emph{glow}): ``I feel like they are doing something because of me and I need to do something as a response.''

P11 (think-aloud, referring to \emph{step}): ``This is a little bit more entertaining, yeah, I think it's more mesmerising to watch if the floor is lighting up.''

\xdef\tpd{\the\prevdepth}
\end{minipage}
\vspace{1em}

We also discovered the influence of metaphors employed on participants' emotions, as five participants noted feeling relaxed or calm through the projection of \emph{wave}, for example, describing it as ``breathing bubbles that reminds me to breathe'' (P20).

\subsection{Perception of Spatial Dynamics}
The eHMIs were designed with varying mechanisms of integration with the physical environment of the shared space. We found that sixteen participants utilised spatial cues presented by these designs to support their interaction with the AV. Many of these cues were not intentional by design but inferred by participants; for example, \emph{step} was often considered to convey the AV's ``planned'' trajectory or its direction of moving, and both \emph{wave} and \emph{step} were perceived useful for marking safe and unsafe areas.

\vspace{1em}
\hfill\begin{minipage}{\dimexpr\textwidth-1cm}

P2 (think-aloud, referring to \emph{glow}): ``I guess it has recognised me...I guess it's a big flashing object so that I can see it a mile away.''

P6 (interview, referring to \emph{wave}): ``The shrinking and growing -- it somehow feels to me it's highlighting the vehicle, as in you can see exactly where the car is located.''

P17 (interview, referring to \emph{step}): ``It can light up the road to show the planned path of the vehicle [and] make sure you don't stand on this path. That's pretty cool.''

\xdef\tpd{\the\prevdepth}
\end{minipage}
\vspace{1em}

Furthermore, since \emph{step} was designed to interact with the paving tiles, six participants had the impression that the AV was an integral part of the place, conveyed through comments like ``It's like the vehicle is there for the purpose to integrate with that space'' (P3), and ``It makes a nice connection with the actual surface of the road -- it seems to make sense for the environment.'' (P14).

\subsection{Effect of eHMIs on Sense of Place}

While appreciating the aesthetic design of these interfaces, eight participants spontaneously identified a resonance with a renowned local light festival, drawing parallels through recollections of similar techniques, such as "lights projected onto a building" (P10) and likening them to specific installations that shared inherent mobile traits with AVs, e.g., ``like the moving lights in a drone show'' (P22). Through comments from nine participants, we found that \emph{step} and \emph{wave} possessed the potential to shape the social dynamics within the shared space, as they were perceived to influence interactivity, both between the AV and pedestrians and among pedestrians themselves:

\vspace{1em}
\hfill\begin{minipage}{\dimexpr\textwidth-1cm}
P13 (think-aloud, referring to \emph{wave}): ``If I was standing here talking to my friends, we might move a little bit but probably continue our conversation and just keep an eye on it, but I wouldn't jump out of the way terrified.''

P14 (interview, referring to \emph{step}): ``It sort of makes me think of Lego or Tetris. I also thought it was the most kind of friendly atmosphere created -- it's the most innocuous, you know, something maybe even kids would like.''

\xdef\tpd{\the\prevdepth}
\end{minipage}
\vspace{1em}

Pertaining to this, four participants speculated on the effect when the paving tile lighting was activated not only by AVs but also by pedestrians or other road users such as cyclists. Contrarily, P7 suggested that the tiles should only light up in response to AV movement, as this would ``distinct that vehicle from any other moving things around'' and ``somehow compensate for the feeling that it's a vehicle without driver''. Furthermore, five participants imagined the designs would allow the AV to perform its utilitarian roles in the space in unconventional or even playful ways, as exemplified by P14's comments on \emph{step}: ``like if you change the colour that [these tiles] would have, it could be just a way to have wayfinding and crowd management in the space in a friendly, fun manner.''

Additionally, seven participants noted the perception that the interfaces appeared to make deliberate efforts to improve their experience within this shared space:

\vspace{1em}
\hfill\begin{minipage}{\dimexpr\textwidth-1cm}
P3 (interview, referring to \emph{step}): ``It's connected with like a sort of a system, where I think that provides an additional layer of, like a sense of safety within that space. At the same time, I would like to trust it because it's something that feels intentional and was designed in a way that was meant to coexist with people around.''

P20 (interview, referring to \emph{wave}): ``It tells me that the goal of the car is not to go somewhere; it's also to be seen and enjoyed by me and be a talking point for me and friends that might be looking at it like, `Oh, how cool'.''

\xdef\tpd{\the\prevdepth}
\end{minipage}
\vspace{1em}

\subsection{Correlations Between Experiential Qualities}
\label{correlation}
\begin{figure}[h]
  \centering
  \includegraphics[width=\textwidth]{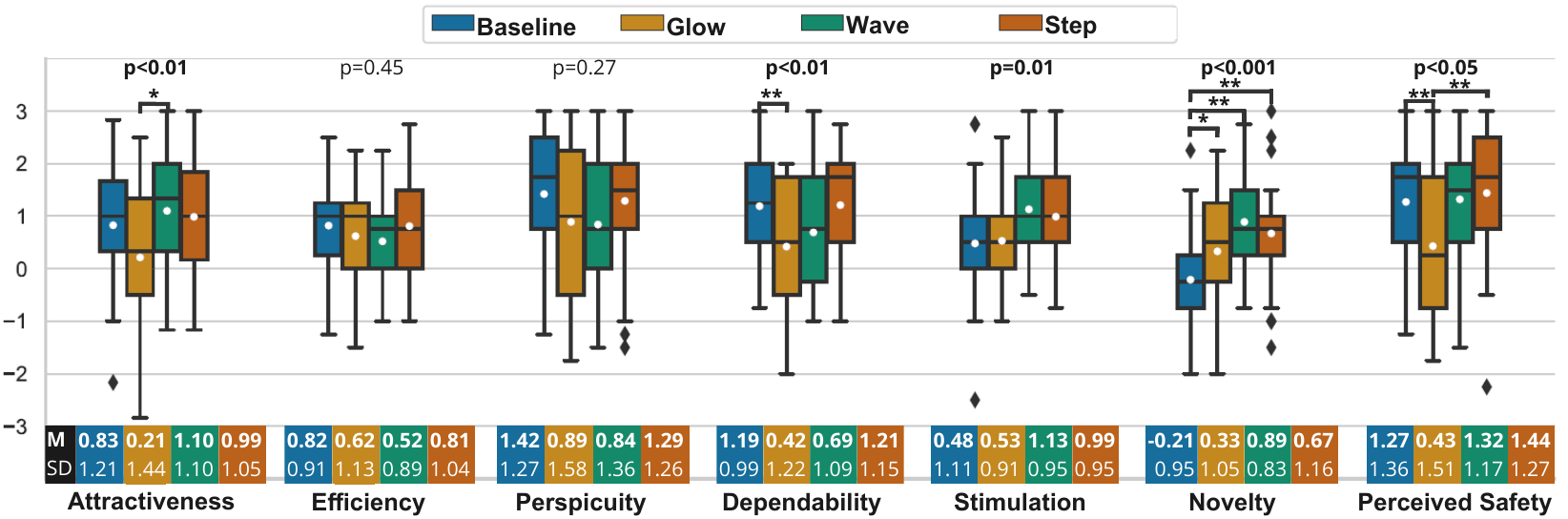}
  \caption{Results of UX and perceived safety. M: mean, SD: standard deviations, *:p<0.05, **:p<0.01. Means are denoted with white circles inside the box plots.}
  \label{ux_safety}
\end{figure}

\begin{figure}[h]
  \centering
  \includegraphics[width=0.42\textwidth]{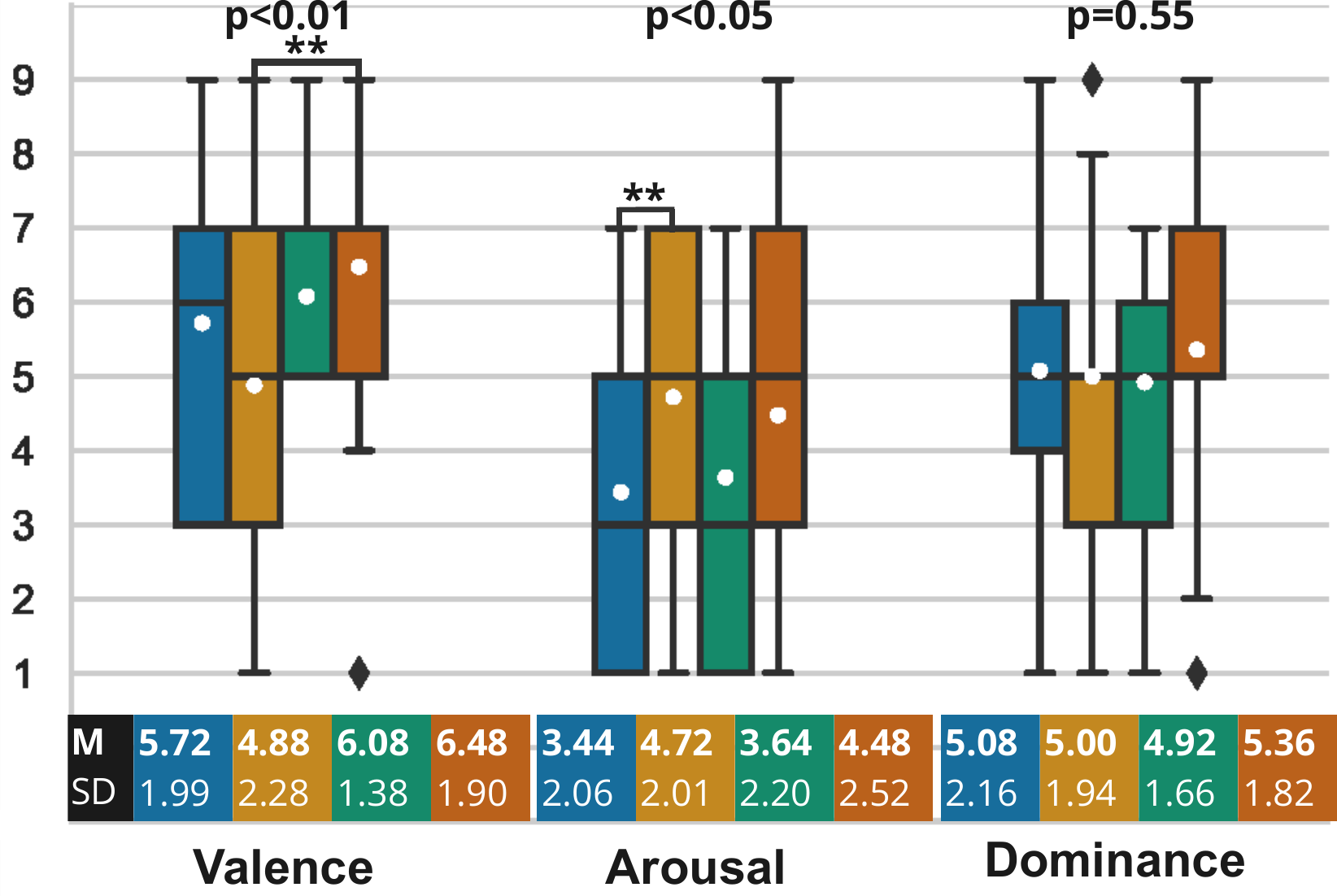}
  \caption{Results of affect. M: mean, SD: standard deviations, *:p<0.05, **:p<0.01. Means are denoted with white circles.}
  \label{emotion}
\end{figure}

\autoref{ux_safety} and \autoref{emotion} illustrates questionnaire responses to UX, perceived safety, and affect for the three place-based eHMI designs as well as the baseline. \autoref{correlation_ux_emotion} and \autoref{correlation_safety} present correlations between these experiential qualities. Table~\ref{ranking} shows participants' preferences towards the four designs. \emph{Step} was the top preference (1st-rank) for nearly half of the participants (n=12). A Friedman test showed a significant difference in the mean rankings among designs ({$\chi^2$}(3) = 51.32, p < 0.001, W = 0.68). Post-hoc tests revealed that \emph{Step} and \emph{Wave} were ranked statistically\footnote{We omit ``statistically'' in later report of statistical significance for brevity purpose.} significantly higher than \emph{Baseline} and \emph{Glow} (all p < 0.001), and \emph{Step} was also ranked significantly higher than \emph{Wave} (p < 0.001). We did not find significant differences between \emph{Glow} and \emph{Baseline} regarding ranking.

\begin{table}[h]
\small
  \caption{Ranking of the four designs (1st-rank=1, 4th-rank=4). M: mean, SD: standard deviations.}
  \label{ranking}
  \begin{tabular}{lcccc}
    \toprule
    \textbf{} & \makecell{\textbf{Baseline} \\ (\textbf{M} / SD)} & \makecell{\textbf{Glow} \\ (\textbf{M} / SD)} & \makecell{\textbf{Wave} \\ (\textbf{M} / SD)} & \makecell{\textbf{Step} \\ (\textbf{M} / SD)}\\
 \midrule
    \textbf{Overall Ranking} & \textbf{2.80} / 1.16 & \textbf{3.04} / 0.98 & \textbf{2.32} / 1.07 & \textbf{1.84} / 0.94 \\
  \bottomrule
  \multicolumn{5}{l}{\footnotesize * A lower ranking indicates higher preference by participants.} \\
\end{tabular}
\end{table}

\begin{table*}[h]
\caption{Spearman's correlation coefficient between affect and UX and perceived safety. Each cell contains four coefficients related to the four eHMIs: baseline, glow, wave, step (from top to bottom). *:p<0.05, **:p<0.01, ***:p<0.001.}
\small
\begin{tabular}{lllllllll}
\hline
\multicolumn{2}{l}{}                                                                                                                & \textbf{\footnotesize Attractiveness}                                                      & \textbf{\footnotesize Efficiency}                                                        & \textbf{\footnotesize Perspicuity}                                                       & \textbf{\footnotesize Dependability}                                                     & \textbf{\footnotesize Stimulation}                                                     & \textbf{\footnotesize Novelty}                                                   & \textbf{\footnotesize Perceived Safety}                                                  \\ \hline
\multicolumn{1}{l}{\textbf{\footnotesize Valence}}   & \textit{\begin{tabular}[c]{@{}l@{}}baseline\\ glow\\ wave\\ step\end{tabular}} & \begin{tabular}[c]{@{}l@{}}0.79***\\ 0.84***\\ 0.53**\\ 0.71***\end{tabular} & \begin{tabular}[c]{@{}l@{}}0.60**\\ 0.68***\\ 0.50*\\ 0.85***\end{tabular} & \begin{tabular}[c]{@{}l@{}}0.57**\\ 0.62***\\ 0.54**\\ 0.56**\end{tabular} & \begin{tabular}[c]{@{}l@{}}0.66***\\ 0.78***\\ 0.42*\\ 0.59**\end{tabular} & \begin{tabular}[c]{@{}l@{}}0.62**\\ 0.56**\\ 0.55**\\ 0.50*\end{tabular} & \begin{tabular}[c]{@{}l@{}}0.01\\ 0.27\\ 0.26\\ 0.19\end{tabular}  & \begin{tabular}[c]{@{}l@{}}0.62***\\ 0.76***\\ 0.37\\ 0.71***\end{tabular} \\ \hline
\multicolumn{1}{l}{\textbf{\footnotesize Arousal}}   & \textit{\begin{tabular}[c]{@{}l@{}}baseline\\ glow\\ wave\\ step\end{tabular}} & \begin{tabular}[c]{@{}l@{}}-0.07\\ -0.28\\ -0.05\\ 0.11\end{tabular}         & \begin{tabular}[c]{@{}l@{}}-0.36\\ -0.31\\ -0.26\\ -0.04\end{tabular}      & \begin{tabular}[c]{@{}l@{}}-0.29\\ -0.33\\ -0.12\\ 0.09\end{tabular}       & \begin{tabular}[c]{@{}l@{}}-0.28\\ -0.50*\\ -0.30\\ -0.39\end{tabular}     & \begin{tabular}[c]{@{}l@{}}0.15\\ -0.09\\ -0.07\\ 0.25\end{tabular}      & \begin{tabular}[c]{@{}l@{}}0.26\\ 0.12\\ 0.16\\ 0.24\end{tabular}  & \begin{tabular}[c]{@{}l@{}}-0.56**\\ -0.42*\\ -0.39\\ -0.42*\end{tabular}  \\ \hline
\multicolumn{1}{l}{\textbf{\footnotesize Dominance}} & \textit{\begin{tabular}[c]{@{}l@{}}baseline\\ glow\\ wave\\ step\end{tabular}} & \begin{tabular}[c]{@{}l@{}}0.46*\\ 0.19\\ 0.25\\ 0.38\end{tabular}           & \begin{tabular}[c]{@{}l@{}}0.47*\\ 0.24\\ 0.14\\ 0.43*\end{tabular}        & \begin{tabular}[c]{@{}l@{}}0.41*\\ 0.39\\ -0.15\\ 0.18\end{tabular}        & \begin{tabular}[c]{@{}l@{}}0.60**\\ 0.40*\\ 0.17\\ 0.52**\end{tabular}     & \begin{tabular}[c]{@{}l@{}}0.29\\ 0.17\\ 0.08\\ 0.23\end{tabular}        & \begin{tabular}[c]{@{}l@{}}-0.02\\ 0.05\\ 0.22\\ 0.24\end{tabular} & \begin{tabular}[c]{@{}l@{}}0.71***\\ 0.28\\ 0.20\\ 0.30\end{tabular}       \\ \hline
\end{tabular}
\label{correlation_ux_emotion}
\end{table*}

\begin{table*}[h]
\caption{Spearman's correlation coefficient between perceived safety and UX. Each cell contains four coefficients related to the four eHMIs: baseline, glow, wave, step (from top to bottom). *:p<0.05, **:p<0.01, ***:p<0.001.}
\small
\begin{tabular}{llllllll}
\hline
\multicolumn{2}{l}{}                                                                                                                                                                 & \textbf{\footnotesize{Attractiveness}}                                                     & \textbf{\footnotesize{Efficiency}}                                                       & \textbf{\footnotesize{Perspicuity}}                                                        & \textbf{\footnotesize{Dependability}}                                                        & \textbf{\footnotesize{Stimulation}}                                                   & \textbf{\footnotesize{Novelty}}                                                   \\ \hline
\multicolumn{1}{l}{\textbf{\begin{tabular}[c]{@{}l@{}}\footnotesize{Perceived}\\ \footnotesize{Safety}\end{tabular}}} & \textit{\begin{tabular}[c]{@{}l@{}}baseline\\ glow\\ wave\\ step\end{tabular}} & \begin{tabular}[c]{@{}l@{}}0.62***\\ 0.82***\\ 0.64***\\ 0.50*\end{tabular} & \begin{tabular}[c]{@{}l@{}}0.65***\\ 0.59**\\ 0.54**\\ 0.53**\end{tabular} & \begin{tabular}[c]{@{}l@{}}0.63***\\ 0.55**\\ 0.64***\\ 0.61**\end{tabular} & \begin{tabular}[c]{@{}l@{}}0.77***\\ 0.74***\\ 0.77***\\ 0.67***\end{tabular} & \begin{tabular}[c]{@{}l@{}}0.40*\\ 0.57**\\ 0.56**\\ 0.25\end{tabular} & \begin{tabular}[c]{@{}l@{}}-0.02\\ 0.27\\ 0.00\\ 0.01\end{tabular} \\ \hline
\end{tabular}
\label{correlation_safety}
\end{table*}

\subsubsection{User Experience}
The designs were considered significantly different in terms of the overall attractiveness (F(3, 72) = 4.27, p < 0.01, $\eta^2 = 0.15$), both of hedonic qualities: stimulation (F(3, 72) = 3.81, p = 0.01, $\eta^2 = 0.14$) and novelty (F(3, 72) = 7.75, p < 0.001, $\eta^2 = 0.24$), and one of the pragmatic qualities: dependability ({$\chi^2$}(3) = 11.99, p < 0.01, W = 0.16). We did not find statistical significance in the other two pragmatic qualities: efficiency (F(3, 72) = 0.90, p = 0.45, $\eta^2 = 0.04$) and perspicuity ({$\chi^2$}(3) = 3.95, p = 0.27, W=0.05). From post-hoc tests, we found that \emph{wave} was rated significantly higher than \emph{glow} for attractiveness (p=0.035). \emph{Baseline} was rated significantly lower in terms of novelty than \emph{glow} (p=0.047), \emph{wave} (p=0.003), and \emph{step} (p=0.006), but rated significantly higher than \emph{glow} for dependability (p=0.001). There were no significant pairwise differences for stimulation.

\subsubsection{Affect}
We found the designs had significantly different effects on participants' emotional valence ({$\chi^2$}(3) = 14.53, p < 0.01, W = 0.19) and arousal ({$\chi^2$}(3) = 9.19, p < 0.05, W = 0.12). Post-hoc tests revealed that \emph{step} induced significantly more positive valence than \emph{glow} (p=0.003). \emph{Glow} led to significantly higher arousal than \emph{baseline} (p=0.002).

From Spearman's correlation coefficient (\autoref{correlation_ux_emotion}), we found a significantly positive correlation between valence and most of UX qualities for all designs. For \emph{glow}, dependability significantly negatively correlated with arousal while positively correlated with dominance. We also found most of the pragmatic qualities significantly positively correlated with dominance for \emph{baseline} and \emph{step}.

\subsubsection{Perceived Safety}
The designs were significantly different in terms of perceived safety ({$\chi^2$}(3) = 8.32, p < 0.05, W = 0.11). Post-hoc tests revealed that \emph{glow} was rated significantly lower than \emph{baseline} (p=0.004) and \emph{step} (p=0.008).

From Spearman's correlation coefficient (\autoref{correlation_ux_emotion} and~\ref{correlation_safety}), we found that perceived safety significantly positively correlated with attractiveness (all designs), all pragmatic qualities (all designs), and stimulation (\emph{baseline}, \emph{glow}, \emph{wave}). We also found perceived safety significantly positively correlated with valence (\emph{baseline}, \emph{glow}, \emph{step}) and dominance (\emph{baseline}) and negatively correlated with arousal (\emph{baseline}, \emph{glow}, \emph{step}).

\section{Discussion}
Our goal was to explore how eHMIs designed with placemaking considerations, i.e., place-based eHMIs, can influence the perception and experience of pedestrians in an urban shared space. Using an RtD approach (RQ1), we created three speculative place-based eHMI designs that extended an intent-based eHMI and demonstrated physical and content integration with this shared space. Results of our evaluation study (RQ2) revealed four themes pertinent to how participants attributed characteristics to the AV, how they perceived the AV interaction and the spatial dynamics, and how the designs affected their sense of place. Furthermore, statistically significant correlations were found between qualities of pedestrian experience across perceived safety, UX, and affect.

\subsection{Implications of Place-Based eHMIs on Pedestrian Perception and Experience}
\subsubsection{Feeling Hedonic, Feeling Safer} In regular road settings, eHMIs are often designed to enhance pedestrian perceived safety through intuitive and efficient communication of intentions, for example, in turn enabling pedestrians to cross streets more swiftly and satisfactorily~\cite{chang2022can,colley2022effects,faas2020longitudinal}. However, the goals of pedestrians in shared spaces are not necessarily to complete transportation tasks quickly but often to engage in various non-transport related activities, such as leisurely strolls or utilising the design of the space for social gatherings~\cite{karndacharuk2013analysis,wang2022pedestrian,predhumeau2021pedestrian}. Given this differentiation in road usage, the external interaction of AVs may need to align more closely with how pedestrians use spaces and their opportunistic goals~\cite[p. 60]{tomitsch2017making}. Through evaluating place-based eHMIs, we found that~\emph{in shared spaces, creating attractive and stimulating eHMIs is also likely to enhance pedestrians' perceived safety and pleasantness, in addition to solely supporting pragmatic qualities}. Over half of the participants found our eHMI designs to be fun and intriguing, attributing lively associations to the AV, such as vitality and the potential to entertain. Indeed,  \emph{step} and \emph{wave} received notably high ratings in terms of hedonic qualities and overall attractiveness. Furthermore, perceived safety and pleasantness were not only positively correlated with the eHMIs' utility but also with their stimulation and overall attractiveness.

\subsubsection{Performing is Caring} Recent deployments of AVs in urban environments have reported challenges they face when entering shared spaces. For example, pedestrian can demonstrate discomfort or interfering behaviours when in close proximity to AVs~\cite{eden2017road,rodriguez2017safety}; AVs can interrupt recreational activities and cause pedestrians to give way~\cite{wang2022pedestrian,predhumeau2021pedestrian}; and pedestrians can feel they have higher road priority over AVs~\cite{merat2018externally,rodriguez2017safety}. Our results revealed that~\emph{place-based eHMI designs conveyed a sense of intentionality behind their creations, as participants inferred thoughtfulness from the expressions and recognised the efforts made to ``coexist with people around''}. One-third of the participants naturally likened the performative values of designs to experiencing interactive installations at an urban light festival and speculated on how they could support or foster social dynamics within this place. Notably, our approach to physical integration assisted the majority of participants to infer safety cues from the spatial dynamics. Particularly, \emph{step}, which exhibited its physical integration through the paving tiles, was rationalised by a quarter of the participants to be an integral part of the locale.

\subsubsection{Perceived Intent Influences Emotion} In our study, we observed that nearly one-third of the participants interpreted the intent of \emph{glow} inversely, thinking the AV was requesting a right-of-way rather than signalling its intention to yield. Associations drawn by participants indicated that~\emph{such a dichotomy might have emerged from the established encoding between strong luminous effects and how the sense of urgency is represented in traffic situations}, such as seeing an~\emph{``ambulance''} or a~\emph{``police siren''}, especially when the participant was in the vehicle's way. In line with this finding was the statistical outcome, as we found participants felt more excited (i.e., arousal) and at the same time less in control (i.e., dominance) when they encountered \emph{glow}. Furthermore, previous research identified that discrepancies between the perceived intent of an AV and the users' expectations, e.g., perceived malfunctions of eHMIs~\cite{m2021calibrating}, can lead to negative emotions such as confusion, frustration, and a sense of insecurity. Hence, interpreting an intent contrary to the prescribed one may have led participants to rate~\emph{glow} as less pleasant (i.e., valence) compared to the other designs.

\subsection{Towards AVs that Support Placemaking Goals of Shared Spaces}
\subsubsection{Walking as Experiencing} With policies reprioritising pedestrian mobility in cities, walking has been considered not merely as a ``mobility turn''~\cite{lynch1995city} but as ways in which cities are experienced, a transitory ``floating life'' to engage with the material environment~\cite{jensen2021pedestrians}. In this context, urban walking environments evolve towards inspiring interest for pedestrians, advocating not only for the basic dimensions necessary for the practicality of walking~\cite{loo2021walking}, but also for the creation of ``rich and stimulating'' streets that inspire pedestrians' thoughts and imagination~\cite{calvert2015exploration,loo2021walking}. Correspondingly, research studies identified that when pedestrians choose which route to walk on, they value, among others, the safety and comfort of walking~\cite{kasraian2021evaluating,miranda2021desirable}, the visual quality of streetscapes~\cite{taylor2003aesthetic,miranda2021desirable}, and the availability of destinations and activities~\cite{loo2021walking,kasraian2021evaluating}. In pursuing these qualities, the presence and priority of vehicles in relation to walking are being carefully balanced, given the legacy conflict between automobile and pedestrian experience~\cite{calvert2015exploration,taylor2003aesthetic}. Nevertheless, vehicles have been establishing important socialities~\cite{sheller2000city} and innovations for the convenience and accessibility of civic life, such as autonomous shuttles facilitating citizen transit, many of which pass by shared spaces as part of their planned routes~\cite{eden2017road,rodriguez2017safety,hoggenmueller2022designing,merat2018externally}. These developments point to the need to design for the coexistence of future AVs and pedestrians in urban road-sharing areas and to support design concepts that cater to walking sensations and environments.

\subsubsection{eHMIs as Enabling Tool} Our work therefore undertook a place-based exploration to support the experience of pedestrians when their space is shared by AVs. Placemaking is fundamentally concerned with enhancing the quality of places, creating environments that people care about and are drawn to, thereby elevating their sense of place~\cite{wyckoff2014definition}. The concept has traditionally focused on improving the entities that reside within these places. However, with the advent of autonomous mobile systems such as AVs, the scope of placemaking has expanded~\cite{wiethoff2021media,shi2023can}, exemplified by self-moving urban robots being designed as a strategy to improve the social dynamics and creative inspiration of urban public spaces~\cite{hoggenmueller2020stop,fischer2014movable}. Although AVs primarily traverse shared spaces rather than inhabit them, their encounters with people have been found to affect street dwelling experiences~\cite{predhumeau2021pedestrian,wang2022pedestrian,pelikan2024encountering}. Notably, the two concepts that were highly rated in our study (\emph{wave} and \emph{step}) were designed with a higher level of physical integration with the place. Our qualitative findings suggest these designs seem to further assist AV in creating a sense of affiliation to its surroundings. We found that participants appreciated hedonic qualities perceived from these interfaces and viewed the AV not just as a mundane vehicle, but as a component in the space that promoted a sense of place~\cite{gustafson2001meanings}. These findings contribute insights into utilising emerging eHMIs from a place-based perspective for the planning of AVs into pedestrian areas, thereby equipping AVs with the capacity to support and create interesting and meaningful pedestrian encounters as they pass by these places.

\subsection{Limitations and Future Work}
Our exploration primarily focused on a visual perspective. In terms of auditory elements, we incorporated a synthetic engine sound for the AV and the ambient noise from the 360-degree recording of the shared space. It is important to note that placemaking extends beyond visual integration to include other sensory designs, such as sounds associated with a place's atmosphere~\cite{lacey2016sonic}, e.g., recorded natural sounds or music that complements the ambience of the location. Considering multisensory factors also aids in serving more vulnerable users of shared spaces, such as individuals with visual impairments resting or navigating in shared spaces~\cite{hammond2013attitudes}. These cases can require eHMI designs to consider more comforting or atmosphere-conducive alerts for informing users of the AV's presence, in addition to relying solely on utilitarian horn sounds~\cite{wang2022pedestrian,zhang2024external}.

In addition, the perception of a place can be influenced by temporal dynamics, for example, pedestrians may use spaces differently depending on the time of day~\cite{tomitsch2017making}. Our evaluation was conducted in a peaceful, early-evening setting. However, it is worth considering whether and how the designs could affect pedestrians' perceptions and experiences at other times differently, and the visibility of these designs may be less effective due to lighting conditions. Future iterations of these designs can explore enhancements such as high-contrast visual elements or dynamic displays that adjust based on ambient lighting. Future research on place-based eHMIs should account for the characteristics of a place at specific times or how perceptions of a design might vary across different times of the day.

Finally, the design proposals may face challenges when transitioning from the VR environment to real-world applications. From a technical perspective, these concepts are grounded in the promising potential of technologies that are gradually maturing and are, to some extent, speculative. As a result, current technologies are not yet ready to support the full implementation of these concepts in the real world. Furthermore, real-world environments induce more complexities in vehicle operations and interactions, including the variability in weather and road conditions and the intricate dynamics of multi-agent scenarios. Future research should gradually investigate these aspects and improve the knowledge scale for applying place-based eHMIs in real life. That said, the results from our study provide a glimpse and starting point into the advantages of adopting such technologies and interaction mechanisms in the future, particularly in how place-based eHMIs can enhance pedestrians' experiences in shared spaces, and hence provide insights for continued investigations around shared-space vehicle technologies and vehicle-pedestrian interactions in highly urbanised zones.

\section{Conclusion}
Urban shared spaces are often designed by city planners with the objective of placemaking. However, the introduction of AVs can sometimes disrupt how these areas are used and enjoyed by pedestrians. Through an RtD process, we investigated the potential of eHMIs, emerging HCI interfaces facilitating AV-pedestrian interaction, for supporting placemaking within these shared spaces. Drawing on urban HCI literature, we developed three speculative, place-based eHMI designs focusing on the physical and content integration with a local shared space and the alignment with the yielding message of an intent-based eHMI. By conducting an immersive evaluation with mixed-methods data collection, we identified that including placemaking considerations could make the eHMIs appear more attractive and stimulating, which simultaneously made the AV interaction perceived as more safe and pleasant. In addition to inferring safety cues from the spatial dynamics of the designs, our participants recognised the intentionality behind these eHMIs, by likening them to interactive installations at an urban light festival and observing their potential to influence social dynamics. From this place-based exploration, we contribute findings to and open the discourse on how eHMIs may be utilised to support future AVs in enhancing the quality of pedestrian-centric environments and the experience of those who encounter them.

\vspace{2em}
\noindent\textbf{Ethical:} This study has been approved by the Human Research Ethics Committee (HREC) at the University of Sydney with protocol number 2022/518.

\vspace{1em}
\noindent\textbf{Data Availability Statement:} Data collected for this study cannot be made publicly available due to ethical restrictions outlined by the University of Sydney HREC. Reasonable requests to access the data can be discussed with the corresponding author in accordance with ethical guidelines.

\vspace{1em}
\noindent\textbf{Authors Contributions:} All authors contributed to the study conception. Material preparation and interface design were conducted by Yiyuan Wang. The ideation workshop was facilitated by Yiyuan Wang, Martin Tomitsch, and Marius Hoggenmüller. Data collection and analysis were performed by Yiyuan Wang and Wai Yan. The manuscript was primarily written by Yiyuan Wang; one section was authored by Marius Hoggenmüller. All authors reviewed and approved the manuscript.

\vspace{1em}
\noindent\textbf{Funding:} This research was funded by the Australian Research Council through grant number DP200102604 Trust and Safety in Autonomous Mobility Systems: A Human-Centered Approach and DP220102019 Shared-Space Interactions Between People and Autonomous Vehicles.

\vspace{1em}
\noindent\textbf{Competing Interests:} The authors declare no competing interest.

\bibliographystyle{ACM-Reference-Format}
\bibliography{reference}

\appendix

\end{document}